\documentclass[pre,floatfix,twocolumn,superscriptaddress]{revtex4}
\usepackage{graphicx} \usepackage{subfigure}
\usepackage{stdclsdv} \usepackage{array}  \usepackage{amsmath}
\usepackage{amsfonts}
\usepackage{xcolor}
\usepackage{float}

\usepackage{hyperref}
\usepackage{xcolor}
\hypersetup{
    colorlinks,
    linkcolor={red!50!black},
    citecolor={blue!50!black},
    urlcolor={blue!80!black}
}

\newcommand{\Fig}[1]{Fig.~\ref{#1}}
\newcommand{\Figs}[1]{Figs.~\ref{#1}}

\newcommand{\Eq}[1]{Eq.~\ref{#1}}

\usepackage{currfile}
\begin{document}

\title{Fiber plucking by molecular motors yields large emergent
contractility in stiff biopolymer networks}
 
\author{Pierre Ronceray}
\affiliation{Princeton Center for Theoretical Science, Princeton University, Princeton, NJ 08544, USA}
\author{Chase P. Broedersz}
\affiliation{Arnold-Sommerfeld-Center for Theoretical Physics and Center for NanoScience, Ludwig-Maximilians-Universit\"at M\"unchen, D-80333 M\"unchen, Germany}
\author{Martin Lenz}
\affiliation{LPTMS, CNRS, Univ. Paris-Sud, Universit\'e Paris-Saclay, 91405 Orsay, France}
\affiliation{MultiScale  Material  Science  for  Energy  and  Environment,
UMI  3466,  CNRS-MIT,
77  Massachusetts  Avenue,
Cambridge,  Massachusetts  02139,
USA}

\begin{abstract}
The mechanical properties of the cell depend crucially on the
tension of its cytoskeleton, a biopolymer network that is put under stress by active motor proteins. While the fibrous nature of the network is known to strongly affect the transmission of these forces to the cellular scale, our understanding of this process remains incomplete. Here we investigate the transmission of forces through the network at the individual filament level, and show that active forces can be geometrically amplified as a transverse motor-generated force force ``plucks'' the fiber and induces a nonlinear tension. In stiff and densely connnected networks, this tension results in large network-wide tensile stresses that far exceed the expectation drawn from a linear elastic theory. This amplification mechanism competes with a recently characterized network-level amplification due to fiber \emph{buckling}, suggesting that that fiber networks provide several distinct pathways for living systems to amplify their molecular forces.
\end{abstract}

\maketitle

\section{Introduction}

Living organisms use chemical energy to produce the mechanical forces required to move and control their shape. These forces originate at the molecular scale from the power strokes exerted by molecular motors, and are generically transmitted by fiber networks such as the actin cortex, thus resulting in large-scale stresses~\cite{Blanchoin:2014}. To develop a theory for active stress generation, it is thus important to understand both how active forces exerted on an individual filament transmit along this filament to the mesh size level, and are subsequently propagated through the network. Stress generation thus involves force transmission through local one-dimensional objects and higher-dimensional networks.

\begin{figure}[hbt]
  \centering
  \includegraphics[width=\linewidth]{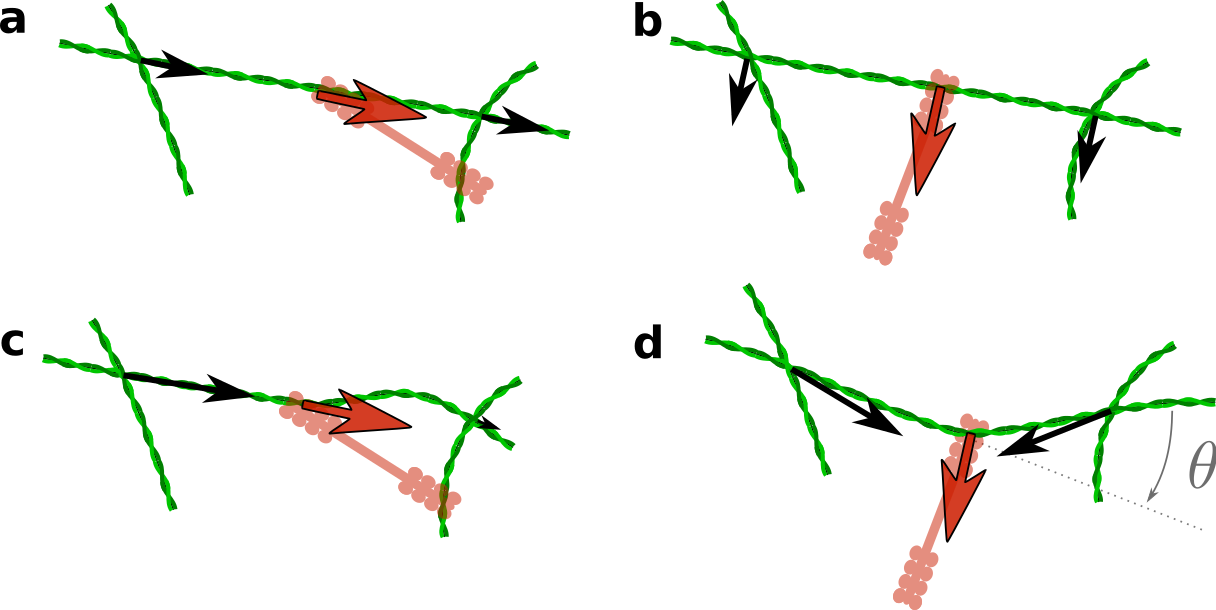}
  \caption{\label{fig:schematics} Local response of actin filaments to myosin forces.
  \textbf{a-b.} Linear response of a  filament to an active force $F$ (\emph{red arrow}) exerted by a motor along and perpendicular to the filament, respectively. Black arrows show the forces transmitted by the filament to the rest of the network.
  \textbf{c.} A large force along the filament leads to buckling of the compressed part. This effectively results in adding a contractile tension $T_\mathrm{buckle}$ in the filament, the magnitude of which is bounded by the motor force ($T_\mathrm{buckle} \lesssim F$).
  \textbf{d.}  A force exerted perpendicularly to the filament tenses it, as does a musician plucking the string of their instrument. Assuming the angular deflection $\theta$ of the filament remains small, this geometrical effect results in a tension $T_\mathrm{pluck}\sim F/\theta$ along the filament's direction which can be much larger than the applied force. The rigidity of the surrounding network is crucial for this amplification mechanism, as $\theta$ can only remain small if the filament is firmly anchored at its ends. 
  }
\end{figure}

Tensile forces propagating through biopolymer networks can be much
longer-ranged than would be expected from linear elastic force
transmission
\cite{Shokef:2012,Notbohm:2014,Rosakis:2014,Xu:2015,Wang:2014},
and have been proposed to enhance cell-cell mechanical communication~
\cite{Sopher:2018}. This suggests that fiber networks do more than
simply transmitting forces: they \emph{amplify}
them~\cite{ronceray_fiber_2016}. This amplification of contractile
stresses can be accounted for by the large-scale mechanical response
of the network surrounding the motors. There, buckling of filaments
under compression increases the range of propagation of tensile
forces, thus enhancing contractile stresses. In order to generate a
significant amplification, this scenario requires active forces that
are much larger than the buckling force of individual filaments, which
induce buckling in a large neighborhood of the motor. As a result, this mechanism for stress generation
is sensitive to the network's mechanical properties and results in
very limited amplification in stiff, densely connected networks.

Here we investigate another stress amplification mechanism that becomes dominant in such stiff networks. This mechanism involves force transmission at smaller scales than the previous one, and we thus consider how forces exerted by a motor on a filament are transmitted by this filament to the rest of the network (\Fig{fig:schematics}\textbf{a}-\textbf{b}). When subjected to a sufficiently large longitudinal motor force, the filament buckles, which at this single-filament level results in a limited force amplification (\Fig{fig:schematics}\textbf{c}). By contrast, a filament under a transverse active force deforms and tenses as a plucked string (\Fig{fig:schematics}\textbf{d}), an effect for which we introduce a single-filament model in Sec.~\ref{sec:model}.
In this geometry, the force transmitted to the mesh size level consists of the initial active force, plus an additional contractile force dipole due to this \emph{plucking} tension. In Sec.~\ref{sec:regimes}, we show that this plucking-induced contractility can become much larger than the one naively expected from the magnitude of the motor force, thus leading to force amplification at the level of a single filament. This geometrical effect is controlled by the deflection angle of the filament, which in turn depends on the stiffness of both
the filament and the network: the stiffer the network, the smaller the deflection, the larger the amplification. By explicitly modeling fiber networks involving many filaments, we show in Sec.~\ref{sec:networks} that this results in a strong dependence of the amplification on the network connectivity, which itself controls its stiffness~\cite{broedersz_criticality_2011}. This study complements our understanding of network-level force amplification to provide a comprehensive theory for force amplification in biopolymer networks.

\section{\label{sec:model}A single-filament model for fiber plucking}

%\subsection{Model}
%\label{sec:model}

\begin{figure}[t] \centering
  \includegraphics[width=\linewidth]{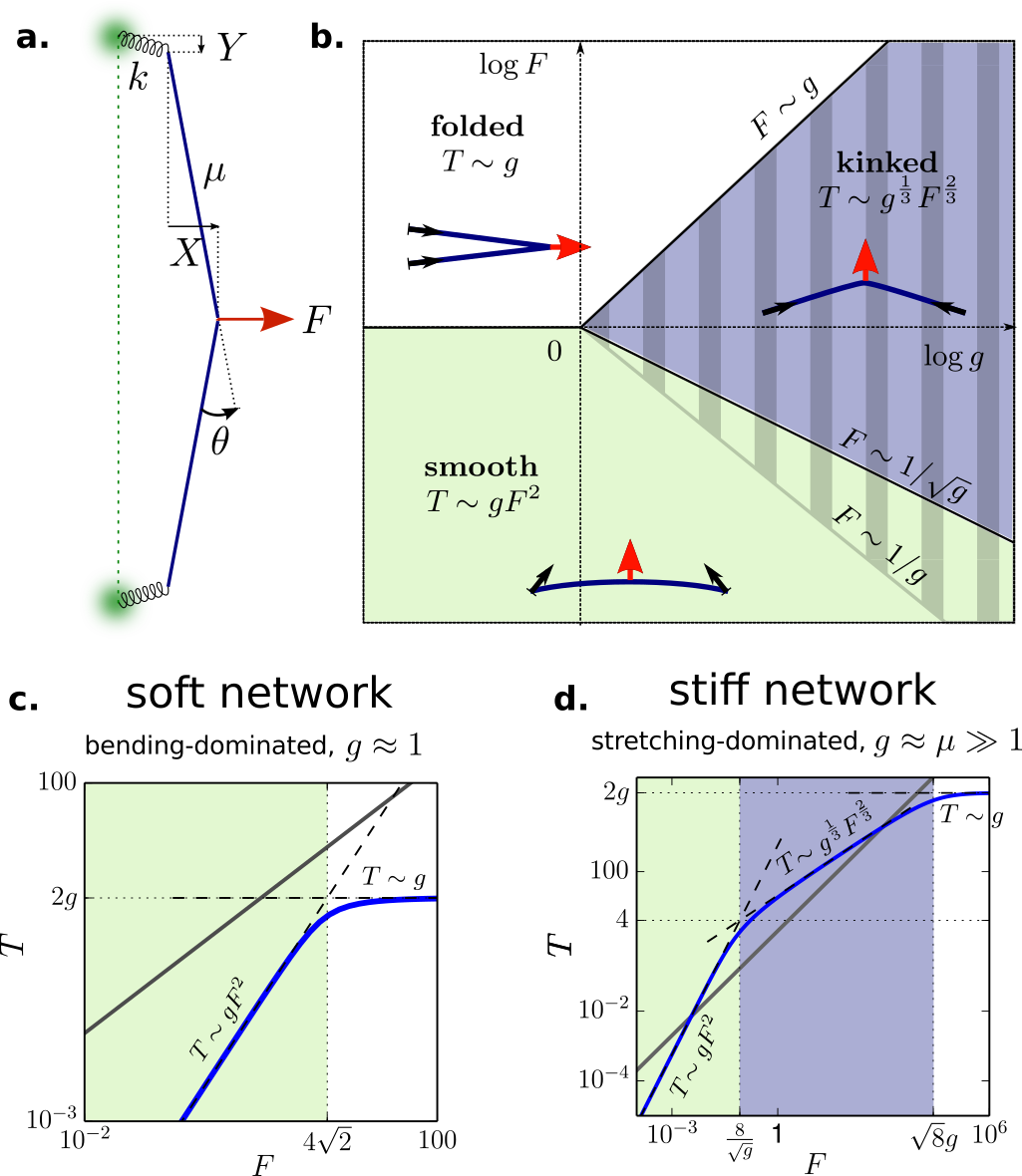}
  \caption{\label{fig:plucking-PD} Single-filament model for filament
    plucking.  \textbf{a.} Model notation, as introduced in
    \Eq{eq:Epluck_nonsimplified}.  \textbf{b.} Scaling regimes for the
    nonlinear tension $T$ that a plucked filament transmits to the
    surrounding network, as a function of the plucking force $F$ and
    the effective rigidity $g$ of the system defined in
    \Eq{eq:g}. Filled areas indicate different scaling regions: smooth
    (green), kinked (blue) and folded (white), with inset depicting
    each situation. The hatched area represents the domain of
    parameters where the plucking tension $T$ exceeds the
    applied force $F$, indicating that plucking dominates the forces
    transmitted to the network.  \textbf{c.} The plucking tension
    (blue line, obtained by minimizing the energy in
    \Eq{eq:Epluck_nonsimplified}) as a function of the active force
    for a soft network with $g=1$; this tension never exceeds the
    applied force $F$ (gray line), and the kinked regime does not
    appear.  \textbf{d.} In a stiff network, $k\sim g\gg 1$, all three
    regimes are present, with a wide regime of force amplification
    where the blue line is above the grey line.}
\end{figure}

To account for the large tensions induced in a filament as a ``plucking'' force is exerted perpendicular to it, we introduce a single-filament model illustrated in
\Fig{fig:plucking-PD}\textbf{a}. The plucked fiber consists of two hinged rods acting under extension as Hookean springs with unit rest length and spring constant $\mu$. An active force $F$ is exerted perpendicular to the filament at the hinge. We penalize any bending of the filament by an angle $\theta$ with an energy $2\sin^2(\theta/2)$. The exact form of this penalty is unimportant as long as it goes as $\theta^2/2$ for small angles, which implies that we set the filament's bending modulus to unity. The surrounding network is modeled as two additional zero-length Hookean springs with spring constant $k$ anchoring the filament to a fixed substrate. The state of the filament is thus represented by two geometrical coordinates $X$ and $Y$, which respectively characterize the filament bending deformation and the network deformation. The energy of the system reads:
\begin{multline}
  \label{eq:Epluck_nonsimplified}
  E = \underbrace{-F X}_{\substack{\text{active}\\ \text{force}}} +
  \underbrace{\left[ \mu (\delta \ell)^2 + 2\sin^2\frac{\theta}{2}
    \right]}_\mathrm{filament\ deformation} + \underbrace{k
    Y^2}_{\substack{\text{network}\\ \text{response}}} 
\end{multline}
with $\delta\ell = \sqrt{X^2 + (1-Y)^2}$ and $\tan(\theta/2)=X/(1-Y)$. In practice, actin filaments are much easier to bend than to stretch, implying $\mu\gg 1$. 

\section{\label{sec:regimes}Force amplification at the single filament level}
To elucidate the magnitude of the contraction generated by our single-filament model, we compute the tensile force $T=kY$ that it transmits to the rest of the network. We first consider the regime of small displacements ($X,Y\ll 1$). Minimizing over $Y$, we obtain the following force balance equation for $X$: 
\begin{equation}
  \label{eq:pluck_X}
F = \frac{\mu k}{\mu+k}  X^3  + X,
\end{equation}
where we omit numerical prefactors for legibility (their precise values give rise to the numerical factors on the
axes of \Figs{fig:plucking-PD}\textbf{c},\textbf{d}). The role of the filament
stretching modulus $\mu$ and the network stiffness $k$ are combined in a
single parameter
 \begin{equation}
  \label{eq:g}
  g = \frac{\mu k}{\mu+k} \approx \min(k,\mu) %\underset{k\ll\mu\text{ or }k\gg\mu}{\sim}
\end{equation}
which corresponds to an \emph{effective stiffness} of the system,
\emph{i.e.} that associated with the softest mode of deformation, either that of the network or that of the filament. Due to the left-right symmetry of \Fig{fig:plucking-PD}\textbf{a}, the tension in the filament is even in the force, and thus has the nonlinear dependence $T \approx gX^2$ in the small elongation limit $\delta \ell \ll 1$.  We can thus rewrite \Eq{eq:pluck_X} in terms of the tension :
\begin{equation}
  \label{eq:pluck_D}
{g}^{1/2} F =   T^{1/2}(1+ T)
\end{equation}
which relates the plucking tension to the applied force and the
mechanical properties of the system. As a result, in the small-displacement regime the value of the plucking tension  depends on the system parameters only through the combination $g^{1/2} F$.

Now considering both small and large displacements, we distinguish between the three amplification regimes illustrated in \Fig{fig:plucking-PD}\textbf{b}, the first two being directly described by \Eq{eq:pluck_D}:
\begin{itemize}
\item at small forces $ F \ll g^{-1/2}$, the filament's response is
  dominated by its bending stiffness, and
  \begin{equation}
T \approx gF^2\label{eq:smooth}
\end{equation}
In this regime, the filament is bent smoothly. If our discrete filament mode were replaced by a continuum one, the same scaling would be obtained and the curvature would be spread over the whole length of the filament.
\item at intermediate forces $ g^{-1/2} \ll F \ll g$, the response is
  dominated by the stretching properties of the filament, and
  \begin{equation}
    T \approx g^{1/3}F^{2/3}\label{eq:kinked}
  \end{equation}
  note that this non-trivial scaling regime only occurs if $g\gg 1$,
  \emph{i.e.} for a stiff filament in a stiff network. The filament is kinked at the hinge in this regime. In a continuum filament with a stretching modulus determined by entropic elasticity, this kink would manifest itself as a localization of the curvature in a region of size $\propto (k_BT/F)^{2/3}\ell_p^{1/3}$ around the point of application of the force, where $\ell_p$ is the filament persistence length~\cite{lenz_geometrical_2014}.
  \item at very large force (or for a very soft network) $F\gg g$, the
  filament folds around its hinge and bends completely; in that
  case the plucking tension saturates to a finite value
  \begin{equation}
    T \approx g\label{eq:folded}
  \end{equation}
  that is independent of the force $F$.
\end{itemize}

To determine whether the effects considered here generate force amplification, we compare the magnitude of the tension $T$ to that of the force $F$ applied by the motor. Amplification occurs in both the smooth and kinked regimes, and specifically for $1/g \ll F \ll g$ as represented by the hatched area in \Fig{fig:plucking-PD}\textbf{b}. The {force amplification} ratio $T/F$ reaches its maximum $\sqrt{g}$ at the crossover between these two regimes. At this crossover $T = 1$, meaning that the tension equals the longitudinal force required to buckle the filament.

\section{\label{sec:networks}Transmission of plucking forces through full networks}

While we have introduced the network rigidity $k$ as independent from the characteristics of the filament of interest, the filaments constituting the network itself are of the same nature as that filament, which strongly constrains the value of the rigidity parameter $g$. There is some freedom in setting the value of $g$ however, as the elasticity of fiber networks is not only controlled by the rigidity of its components, but also by its architecture, and in particular its connectivity. Indeed, densely connected networks primarily deform through stretching modes and have $k\sim \mu$, while loosely connected networks deform through bending modes, and thus have $k\sim 1$ in our units~\cite{broedersz_modeling_2014}. Here we investigate how these different architectures impact force amplification through plucking.

Going back to our single-filament model, we predict that in soft bending-dominated networks with $k\sim 1$ (and thus $g\sim 1$), the plucking tension is small compared to the applied force, regardless of the value of $F$ (\Fig{fig:plucking-PD}\textbf{c}). In contrast, for stretching-dominated networks with $k\sim \mu$ (and thus $g\sim\mu \gg 1$), there is a vast regime of force values for which amplification occurs (\Fig{fig:plucking-PD}\textbf{d}).
  
\begin{figure}[t] \centering
  \includegraphics[width=\linewidth]{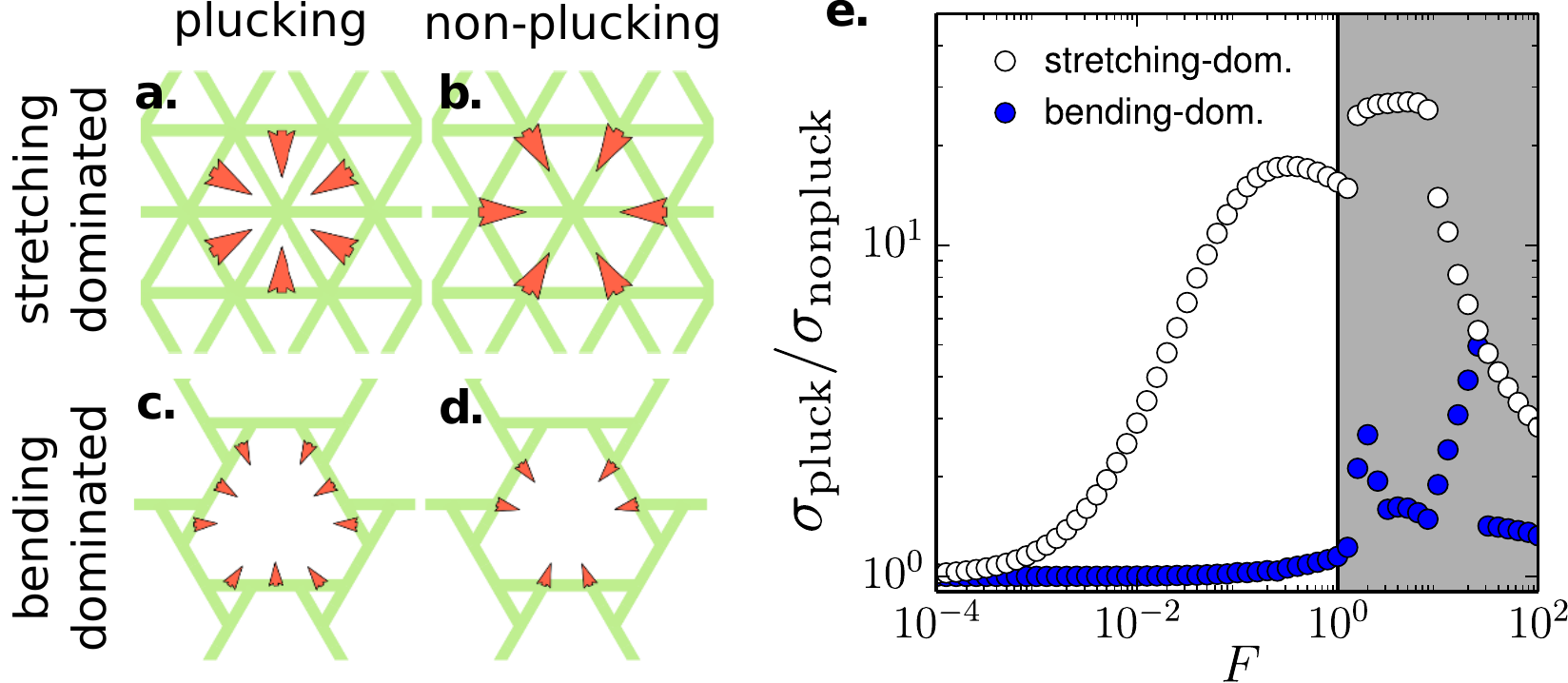}
 \includegraphics[width=\linewidth]{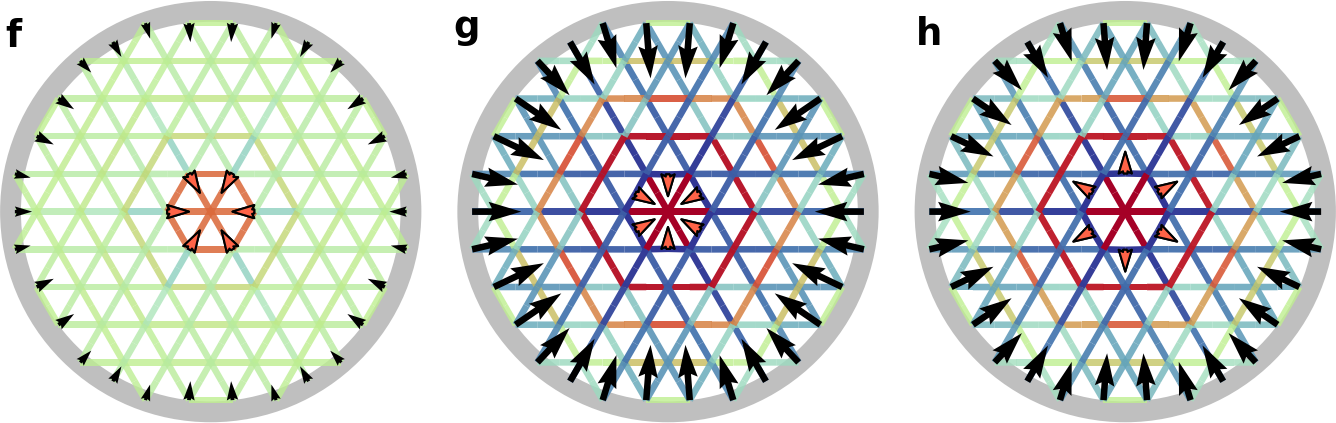}
 \caption{\label{fig:amplification}Full-network model of
   plucking. \textbf{a-d.} Local force application geometries.
\textbf{e.} Ratio of the stresses generated in the presence \emph{vs.} absence of plucking in a circular stretching- (bending-)dominated network with radius $R=11$. In the gray area buckling occurs, resulting in large displacements and a complex network response involving the effects described in Ref.~\citep{ronceray_fiber_2016}.
\textbf{f-g.} Force propagation in a stretching-dominated network in the non-plucking and plucking geometries, respectively, at force $F=0.2$. Blue (red) bonds are under tension (compression). Black arrows indicate forces at a fixed boundary.
\textbf{h.} Reversing forces compared to the previous panel leaves the stress pattern almost unchanged. Active force and stress scale are the same in panels {f-h}. Boundary forces are magnified 5 times in panel {f.} for legibility. In all simulations $\mu=10^4$.}
\end{figure}

To put this simple picture to the test, we simulate full networks
using a numerical model where filaments are positioned at the edges of
a regular two-dimensional lattice. Similar to the model of
\Fig{fig:plucking-PD}\textbf{a}, here all filaments are modeled by
hinged Hookean springs of unit rest length, with a stretching energy
$\mu (\delta\ell)^2/2$ associated with an elongation $\delta\ell$ and
a bending penalty $2\sin^2(\theta/2)$ for creating an angular
deflection $\theta$ between two consecutive, initially aligned
springs.  In addition to their hinges at the lattice nodes, filaments
have an additional hinge in between lattice nodes
similar to that pictured in \Fig{fig:plucking-PD}\textbf{a}, implying that they buckle if
subjected to a longitudinal compressive force larger than the critical
force $F_\mathrm{buckling} = 1$. Distinct filaments are connected at each lattice vertex by crosslinks that do not constrain their relative angles. The energy is minimized with respect to
the position of all nodes using the BFGS algorithm. The critical
connectivity delimiting the bending- and stretching dominated regimes
discussed above can be estimated through a simple constraint counting
argument, which indicates that networks where each vertex has more
than $z=4$ neighbors in two dimensions are stretching-dominated, where
those with lower connectivity are
bending-dominated~\cite{broedersz_criticality_2011}. Here we simulate one
network characteristic of each regime: a high-connectivity ($z=6$),
stretching-dominated triangular lattice
(\Fig{fig:amplification}\textbf{a},\textbf{b}), and a low-connectivity
($z=3$), bending-dominated network illustrated in
\Fig{fig:amplification}\textbf{c},\textbf{d} and further described in
Appendix~\ref{sec:MAN}.

To quantify the amount of contractile tension induced by a single motor in a circular piece of network with fixed boundaries, we measure the coarse-grained macroscopic contractile stress (corresponding to the ``active stress'' of active gel theories~\cite{Kruse:2005aa,Prost:2015}), which reads~\cite{Ronceray:2015}
\begin{equation} 
\sigma= -\frac{1}{V}\sum_{i} \mathbf{f}_i \cdot \mathbf{r}_i\label{eq:Dfar}
\end{equation}
where the sum runs over the vertices that belong to the boundary of the network, and $\mathbf{f}_i$ is the force exerted on the boundary at node $i$ located at position $\mathbf{r}_i$. To assess the contribution of plucking to this contractile stress, we compare situations with and without plucking. The former situation is illustrated in \Figs{fig:amplification}\textbf{a},\textbf{c}, the motor forces are exerted at the midpoint of each filament. The latter corresponds to forces being applied to the lattice vertices as shown in \Figs{fig:amplification}\textbf{b},\textbf{d}. We plot the ratio $\sigma_\text{pluck}/\sigma_\text{nonpluck}$ of contractile stresses in the two situations in \Fig{fig:amplification}\textbf{e}. For bending-dominated networks, it is roughly equal to one for all force values, indicating that plucking is of little relevance in these systems, as predicted by our local calculation (\Fig{fig:plucking-PD}\textbf{c}). In contrast, in
stretching-dominated networks the plucking contribution is large, with a stress ratio culminating at a large value $\approx 17$. This value is compatible with the expectation drawn from our one-filament model, which predicts an amplification of the order of $\sqrt{g}$, with $\sqrt{g}\approx \sqrt{\mu}=100$ here.
This strong amplification is evident when comparing \Figs{fig:amplification}\textbf{f} and \textbf{g}, showing the force transmission pattern in the stretching-dominated network with forces
applied in a non-plucking and plucking geometry, respectively. The macroscopic stress of the latter system is 17 times larger than that of the former.

The plucking tension is intrinsically contractile in simple networks, as plucking a filament always tends to shorten its end-to-end distance. As a result, the far-field plucking stress does not depend on the orientation of the force perpendicular to the filament, but only on its magnitude. We illustrate this by reversing the active forces of \Fig{fig:amplification}\textbf{g} in \Fig{fig:amplification}\textbf{h} without any significant alteration of the force transmission pattern. As a result, plucking results in a reversal of locally extensile forces into far-field contractile forces, reminiscent of the buckling-induced rectification reported in Ref.~\cite{ronceray_fiber_2016}.

\section{Discussion}

Our results show that the generation of large-scale
active stress $\sigma$ in stiff biopolymer networks depends not only
on the magnitude $F$ of the local active forces, but also crucially on
the geometry of force exertion. Forces exerted at filament
intersections or along filaments induce a linear stress response $\sigma \propto F$ if the force is smaller than the filaments' buckling threshold. In contrast, forces exerted perpendicularly to a filament far from filament intersections can induce a strongly nonlinear stress
response even for forces well below this threshold. The tension
resulting from this geometrical effect, which we term \emph{plucking},
can dominate the far-field response, strongly amplify contractile
stresses and reverse locally extensile forces. This plucking
effect is generic and independent of spatial dimension.

While plucking can amplify stresses by over an order of magnitude for biologically relevant parameters, it involves several requirements. First, it requires moderate forces, sufficient to bend the filament that they are applied to but not large enough to fold it or buckle its neighbors (\Fig{fig:amplification}\textbf{e}). Second, plucking necessitates a stiff, densely connected networks of flexible filaments, and is negligible in softer, bending-dominated gels. This point was missed in previous studies of local motor-induced network deformations~\cite{lenz_geometrical_2014}. Beyond the linear networks considered here, stiffening under stress both of individual biopolymers and whole gels could thus enhance stress amplification by plucking.

Plucking is distinct from buckling as a stress amplification mechanism, although they share some characteristics. Indeed, both are nonlinear single-filament effects that rectify all local forces towards contraction while strongly amplifying far-field stresses.
These effects are each characterized by specific signatures that could permit to experimentally assess their individual pertinence.
Plucking amplifies forces at a purely local level, making it highly sensitive to the local geometry of force exertion. The resulting amplification culminates at intermediate force and increases in magnitude with network stiffness. Buckling, in contrast, is essentially insensitive to local geometry. At larger scales however, buckling nonlinearly amplifies stress through an increased range of force transmission -- as in a rope network, rather than an elastic medium. This amplification always increases with increasing force, but decreases with increasing network stiffness as filament buckling becomes more difficult. Despite their differences, buckling and plucking can cooperate: the large local forces emerging from plucking in the kinked regime exceed the buckling threshold, and are thus further amplified by buckling the surrounding
network through force transmission.

Plucking and buckling thus provide distinct routes to amplify
biological forces. Buckling amplifies forces at any strain, allowing cells in the extracellular matrix to enhance stress transmission and modify the mechanical properties of the matrix over large distances~\cite{Han201722619}. Plucking, in contrast, generates large stresses only at small strains. It could thus help build tension in the cell's actomyosin cortex, and has been observed in reconstituted actomyosin networks~\cite{Seara:2018}.

\begin{acknowledgments}
This work was supported by a PCTS fellowship to PR, the German Excellence Initiative via the program ``NanoSystems Initiative   Munich''   (NIM) and   the Deutsche Forschungsgemeinschaft   (DFG)   via   project   B12   within   the   SFB-1032 to CPB, Marie Curie Integration Grant PCIG12-GA-2012-334053, ``Investissements d'Avenir'' LabEx PALM (ANR-10-LABX-0039-PALM), ANR grant ANR-15-CE13-0004-03 and ERC Starting Grant 677532 to ML. ML's group belongs to the CNRS consortium CellTiss.
\end{acknowledgments}

\bibliographystyle{unsrt}

\newpage
\appendix
  
\section{A regular bending-dominated network geometry }
\label{sec:MAN}
\begin{figure}[H]
  \centering
  \includegraphics[width=\linewidth]{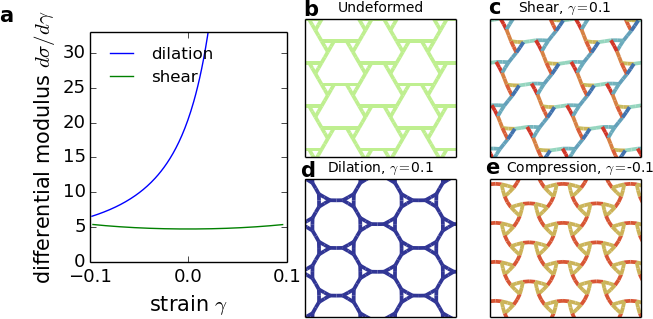}  
  \caption{Elasticity and geometry of a minimal bending-dominated network.
    \textbf{a.}  Macroscopic stress-strain curves under bulk dilation and shear.  \textbf{b.} The network in the
    undeformed configuration. \textbf{c.} Response to shear. \textbf{d.} Response to uniform dilation. \textbf{d.} Response to uniform compression. Red: compressed bonds, blue: tense bonds. Filament stretching modulus $\mu = 10^3$.}
  \label{fig:MAN}
\end{figure}

As described in the main text, fiber networks explore different elastic regimes depending on their average connectivity. While these different regimes have previously been studied in disordered networks~\cite{broedersz_criticality_2011}, here we avoid the technical difficulties associated with disorder by designing a minimal regular, bending-dominated sublattice of the triangular lattice. Its local structure is illustrated in \Fig{fig:amplification}\textbf{c},\textbf{d} of the main text. To further characterize the structure and elasticity of this network, we describe its elastic and geometrical response to shear and dilation strain in \Fig{fig:MAN}. All deformation modes presented show non-affine deformations characteristic of bending-dominated networks. As expected for a bending-dominated network, the network's elastic moduli are of order one, \emph{i.e.}, on par with the filament bending rigidity.
By contrast, the moduli of a highly coordinated, stretching-dominated triangular network are of order $\mu\gg 1$.

\end{document}